\documentstyle[12pt]{article}
\title{{Dynamical Vacuum in Quantum Cosmology}
\author{\sc{Fl\'avio G. Alvarenga and Nivaldo A. Lemos}\\
\small{\it Departamento de F\'{\i}sica}\\
\small{\it Universidade Federal Fluminense}\\
\small{\it Av. Litor\^anea s/n, Boa Viagem }\\
\small{\it 24210-340, Niter\'oi, RJ - Brazil}\\
\small{Electronic mail: flavio@if.uff.br and nivaldo@if.uff.br}}
 }

\begin{document}
\pagestyle{myheadings} 
\baselineskip 22pt
 
\maketitle

\centerline{ABSTRACT}
\vskip .1cm

By regarding the vacuum as a perfect fluid with equation of state $p = - \rho$,
de Sitter's cosmological model is quantized.  
Our treatment differs from previous ones in that it endows the vacuum with  dynamical degress of freedom,
following modern ideas that
the cosmological term is a manifestation of the vacuum energy. 
Instead of being postulated from the start, the cosmological constant  arises  from the degrees of freedom of the vacuum
regarded as a dynamical entity, and a time variable can be naturally introduced. Taking the scale factor  as the sole 
degree of freedom of the gravitational field, stationary  and wave-packet solutions to the
Wheeler-DeWitt equation are found, whose properties are studied. It is found that states of the Universe with  a definite value of the cosmological constant do not exist. For the wave packets
investigated, quantum effects are noticeable only for small values of the scale factor, a classical regime being attained at asymptotically large times.


\newpage
\noindent{\bf 1. INTRODUCTION} 

Quantum cosmology is hopefully relevant to describe quantum gravitational effects in the very early Universe. In view of the nonexistence of a consistent quantum theory of gravity, minisuperspace quantization, which consists in ``freezing out''
all but a finite number of degrees of freedom of the gravitational field and its sources and quantizing the remaining ones,  
is expected to provide general insights on what an acceptable quantum gravity should be like. This line of attack, initiated by DeWitt [1],  has been extensively pursued to quantize model universes with different symmetries and varying matter content, and
 allows one to conceive theories of initial conditions for the wave function of the Universe [2]. Manifold schemes have been devised
to quantize gravity coupled to matter in minisuperspace, the commonest of such quantization methods being those that rely on the Wheeler-DeWitt equation, advocate the quantization of only the conformal factor of the spacetime metric, or perform canonical 
quantization in the reduced phase space.

In inflationary cosmology
de Sitter's model plays a fundamental role, since it describes the phase of rapid expansion during which the vacuum energy dominates the energy density of the Universe, and gives rise to a term in the energy-momentum tensor that corresponds to a cosmologi
cal constant. In modern cosmology the terms {\it vacuum energy} and {\it cosmological constant} are used almost synonymously [3]. It seems, therefore, of interest to study quantum aspects of de Sitter's cosmological model by treating the vacuum as a dynam
ical entity. In such a treatment, the cosmological constant should not be postulated from the start, but should emerge from the dynamical degrees of freedom of the vacuum. A possible way to achieve this is by regarding the vacuum as a perfect fluid with e
quation of state $p=-\rho$. This approach appears to be fruitful,
has several attractive features from the thermodynamic point of view, and leads to interesting consequences  in inflationary cosmology [4]. The standard way of dealing with de Sitter's model in quantum cosmology [5] is highly questionable because it invol
ves  a system with a single degree of freedom and one constraint, so that, strictly speaking, the system has no degrees of freedom at all and is empty of physical content.  The assignation of dynamical degrees of freedom to the vacuum circumvents this dif
ficulty and renders our method distinctive in its ability to make room for the introduction of a time variable.

Accordingly, we shall adopt Schutz's canonical formalism [6]
which  describes a relativistic fluid interacting with the gravitational field. This formalism is especially adequate for our purposes, inasmuch as it has the advantage of ascribing dynamical degrees of freedom to the fluid. As it will be seen, Schutz's a
ction principle  is successful even in the case of the vacuum in the sense that  the cosmological constant appears dynamically as a manifestation of the  degrees of freedom of the fluid that acts as the vacuum. 

In the quantum realm the properties of de Sitter's model will be investigated on the basis of the associated Wheeler-DeWitt equation. Because the super-Hamiltonian constraint is linear in one of the momenta, the Wheeler-DeWitt equation can be reduced to a
 bona fide Schr\"odinger equation. 

This paper is organized as follows. In Section 2 a Hamiltonian treatment of de Sitter's model is developed on the basis of Schutz's canonical formalism, which is proved, in the case of the vacuum, to lead to the correct classical equations of motion. In S
ection 3 the Wheeler-DeWitt equation is written down and  is shown to take the form of a genuine Schr\"{o}dinger equation for an appropriate form of the inner product. In order for the Hamiltonian operator to be self-adjoint its domain must be restricted 
to wave functions that obey certain boundary conditions. General sets of stationary solutions to the Wheeler-DeWitt equation obeying said boundary conditions are found. Then, in Section 4, normalized wave-packet solutions to the Wheeler-DeWitt equation ar
e found, and their properties analyzed. Section 5 is dedicated to final comments.


\vspace{.7cm}

\noindent{\bf 2. DYNAMICAL VACUUM IN DE SITTER'S COSMOLOGICAL MODEL} 

The line element for a homogeneous and 
isotropic universe
can be written in the
Friedmann-Robertson-Walker form (we take $c=1$)

$$ds^2 = g_{\nu\lambda} dx^{\nu} dx^{\lambda} = - N(t)^2 dt^2 
+ R(t)^2 {\sigma}_{ij} dx^i dx^j
\,\,\, , \eqno(2.1)$$                                                  
\\
\noindent where ${\sigma}_{ij}$ denotes the   metric for a 3-space 
of constant curvature
$k= +1, 0$ or $-1$, corresponding to spherical, flat or hyperbolic spacelike
sections, respectively.           

The matter content will be taken to be a perfect fluid, and 
Schutz's canonical formulation of the dynamics of a relativistic fluid in 
interaction with the gravitational field
will be employed [6]. The degrees of freedom ascribed to
the fluid are five scalar potentials $\varphi , \alpha , \beta , \theta , S$ 
in terms of which the  four-velocity of the fluid is written as

$$U_{\nu} = \frac{1}{\mu} \, (\varphi _{,\nu} + \alpha \beta _{,\nu}  +
\theta S_{,\nu}) \,\,\, , \eqno(2.2)$$
\\
\noindent where $\mu$ is the specific enthalpy. By means of the normalization
condition

$$  g_{\nu\lambda} U^{\nu} U^{\lambda} = - 1    \eqno(2.3)$$
\\
one can express $\mu$ in terms of the velocity potentials. The action for the
gravitational field plus perfect fluid is

$$ S= \int_M\,d^4x \sqrt{-g}\,\, ^{(4)}R \, + \, 
2\int_{\partial M}\,d^3x\sqrt{h}\, h_{ij} K^{ij} \, + \, 
\int_M\,d^4x  \sqrt{-g}\,p \,\,\, \eqno(2.4)$$
\\

\noindent in units such that $c=16\pi G=1$. In the above equation
$p$ is the pressure of the fluid, $^{(4)}R$ is the scalar
curvature derived from the spacetime metric $g_{\nu\lambda}$, $h_{ij}$ is the
3-metric on the boundary $\partial M$ of the  4-manifold $M$, 
and $K^{ij}$ is the extrinsic curvature or second fundamental
form of the boundary [7]. The surface term is necessary in the
path-integral formulation of quantum gravity in order to rid the
Einstein-Hilbert Lagrangian of second-order derivatives. Variations of the
pressure are computed from the first law of thermodynamics.

Compatibility with
the homogeneous spacetime metric is guaranteed 
by taking  all of the velocity potentials of the fluid as functions of $t$
only. We shall take $p = (\gamma - 1)\, \rho$ as  equation of state
for the fluid, where $\gamma$ is a constant and $\rho$ is the fluid's energy
density (we shall eventually put $\gamma = 0$).
In the geometry characterized by (2.1) the appropriate 
boundary condition for the
action principle is to fix the initial and final hypersurfaces of constant
time. The second fundamental form of the boundary becomes 
$K_{ij} = - {\dot{h}}_{ij}/2N$. As described
in its full details in [8], after inserting the metric (2.1) into the 
action (2.4), using the equation of state, computing the canonical momenta
and employing the constraint equations to eliminate the pair $(\theta,
p_{\theta})$, what remains is a reduced action in the Hamiltonian form

$$ S_r = \int dt \Bigl \{ {\dot R}p_R + {\dot \varphi}p_{\varphi}
+{\dot S} p_S
- N {\cal H} \Bigr \} \,\,\,  \eqno(2.5)$$                                   
\\
\noindent where an overall factor of the spatial
integral of $(det \, \sigma)^{1/2}$ has been discarded, 
since it has no effect on
the equations of motion. The super-Hamiltonian $\cal H$ is given by 

$${\cal H} = -  \frac{p_R^2}{24R} - 6kR  +
p_{\varphi}^{\gamma}\, R^{-3(\gamma - 1)}\, e^S \,\,\, . \eqno(2.6)$$          
\\
\noindent  The lapse $N$ plays the role of a Lagrange multiplier, 
and upon its variation it is found that the super-Hamiltonian $\cal H$ vanishes.
This is a constraint, revealing that the phase-space 
contains redundant canonical
variables.                                                                     
                               
For $\gamma = 0$ the super-Hamiltonian  contains neither the fluid's degree of
freedom $\varphi$ nor its conjugate momentum $p_{\varphi}$, so that these canonical variables can 
be simply dropped. Equivalently, the correct
classical equations of motion can be obtained without taking into account the 
degrees of freedom described by $\varphi , \alpha , \beta$ and $\theta$, 
that is, they could have been
disregarded from the start. It is a pleasant circumstance that only the physically meaningful entropy density $S$ is
relevant for $\gamma = 0$. The action
reduces to

$$ S = \int dt \Bigl \{ {\dot R}p_R  + {\dot S} p_S
- N {\cal H} \Bigr \} \,\,\,  \eqno(2.7)$$                           
\\

\noindent with

$${\cal H} = -  \frac{p_R^2}{24R} - 6kR  +
 R^3\, e^S \,\,\, . \eqno(2.8)$$                                 
\\

\noindent This can be put in a more suggestive form by means of the canonical
transformation

$$ T = - e^{-S}\, p_S \,\,\,\,\,\,\,\,\,\, , \,\,\,\,\,\,\,\,\,\,
p_T = e^S \,\,\, . \eqno(2.9)$$
\\

\noindent Then 

$$ S = \int dt \Bigl \{ {\dot R}p_R +  {\dot T} p_T
- N {\cal H} \Bigr \} \,\,\,  \eqno(2.10)$$                     
\\                                                           

\noindent where

$${\cal H} = -   \frac{p_R^2}{24R} - 6kR  +
 R^3\, p_T \,\,\, . \eqno(2.11)$$                                 
\\
The extended phase-space is generated by $(R,T,p_R,p_T)$. 
The variable $T$ is such that the Poisson bracket

$$\{ T,{\cal H}\} _{{\cal H}=0} = R^3 >0 \,\,\, , \eqno(2.12)$$
\\
\noindent so that $T$ is a ``global phase time'' or, more precisely, since it
does not involve the canonical momenta, a ``global time'' in accordance with
the terminology introduced by
Hajicek [9]. 
 This is reassuring because the existence of a global time appears to
be a necessary condition to prevent violations of unitarity in the quantum
domain.

The classical
equations of motion are

$$ {\dot R} = \frac{\partial (N {\cal H})}{\partial p_R} = - \frac{N p_R}{12 R}
\,\,\, , \eqno(2.13a)$$
\\                                                                             
$$ {\dot p}_R = -\frac{\partial (N {\cal H})}{\partial R} = 
N \Bigl (- \frac{ p_R^2}{24 R^2} +6k - 3 R^2\, p_T \Bigr )
\,\,\, , \eqno(2.13b)$$
\\                                                                             
$$ {\dot T} = \frac{\partial (N {\cal H})}{\partial p_T} = N R^3
\,\,\, , \eqno(2.13c)$$
\\                                                                             
$$ {\dot p}_T =- \frac{\partial (N {\cal H})}{\partial T} = 0
\,\,\, , \eqno(2.13d)$$
\\                                                                             
\noindent supplemented by the super-Hamiltonian constraint

$$ {\cal H} = - \frac{p_R^2}{24R} - 6kR  +
 R^3\, p_T  = 0 \,\,\, . \eqno(2.14)$$                                 
\\                                                 

In order to solve these equations in the case $k=0$ let us choose the gauge 
$t = T$, so that
$ N=R^{-3}$. Since $p_T $ is constant, it follows from ${\cal H} = 0$ that
$p_R$ is proportional to $R^2$. Insertion of this result in Eq.(2.13a) leads to 

$$R^2 {\dot R} = constant  \,\,\, \Longrightarrow \,\,\, R(t) = (At)^{1/3}
\,\,\, , \eqno(2.15)$$
\\
\noindent where $A$ is a positive constant and the origin of the time $t$ has been conveniently chosen. The lapse function is,
therefore,

$$ N(t) = R^{-3} = (At)^{-1} \,\,\, . \eqno(2.16)$$
\\

\noindent In terms of the cosmic time $\tau$ defined by

$$d\tau = N(t)\, dt = \frac{dt}{At} \,\,\, \Longrightarrow   
\,\,\, \tau - {\tau}_0  = \frac{\ln (t)}{A} \eqno(2.17)$$
\\

\noindent one recovers  the usual de Sitter solution

$$ R = R_0 \, e^{\Lambda \, \tau} \,\,\, . \eqno(2.18)$$
\\
                                     
\noindent This concludes the verification that the action (2.10) leads to de
Sitter's spacetime solution to Einstein's equations. Note that if the time variable is chosen as $t=T$
the scale factor vanishes at $t=0$, whereas in the cosmic-time gauge it vanishes only  at $\tau = -\infty$. It is also worth mentioning that Eqs.(2.9) and (2.13d) show that the entropy density $S$ remains constant, in agreement with the behavior of inflat
ionary models during the de Sitter phase [3].

The general case of arbitrary $k$ can be easily handled in the gauge $N=1$ and
leads to the expected solutions, but we shall refrain from considering it here.



\vspace{.7cm}

\noindent{\bf 3.  QUANTIZED MODEL: A  WHEELER-DeWITT DESCRIPTION}        

It will be
convenient to introduce a new parametrization of the lapse function by
writing it as $NR$. Then the action retains the form (2.10) but the
super-Hamiltonian is now

$$ {\cal H} = - \frac{p_R^2}{24} - 6kR^2  +
 R^4\, p_T  = 0 \,\,\, . \eqno(3.1)$$                                 
\\                                                 
 
\noindent The Wheeler-DeWitt quantization scheme consists in setting

$$ p_R \rightarrow -i\frac{\partial}{\partial R} \,\,\,\,\,\,\,\,\,\, ,
\,\,\,\,\,\,\,\,\,\, p_T \rightarrow -i\frac{\partial}{\partial T} \eqno(3.2)$$
\\
\noindent to form the operator ${\hat {\cal H}}$, and
imposing the Wheeler-DeWitt equation

$$ {\hat{\cal H}} \, \Psi = 0 \eqno(3.3)$$
\\
\noindent on the wave function of the universe $\Psi$. In our present case this
equation takes the form

$$ \frac{1}{24} \frac{\partial ^2 \Psi}{\partial R^2} -6kR^2\, \Psi -iR^4
\frac{\partial \Psi}{\partial T} = 0  
\,\,\, . \eqno(3.4)$$                         
\\

Upon division by $R^4$ this equation takes the form of a Schr\"odinger
equation

$$i\frac{\partial \Psi}{\partial T} =\frac{1}{24R^4} \frac{\partial ^2 \Psi}{\partial R^2} -\frac{6k}{R^2} \eqno(3.5)$$
\\
\noindent with $T$ playing the role of time. In order to be able to interpret $T$ as a true time and (3.5) as a genuine Schr\"odinger equation, the operator

$${\hat H} = \frac{1}{24R^4} \frac{\partial ^2 }{\partial R^2} -\frac{6k}{R^2}\eqno(3.6)$$
\\
\noindent must be self-adjoint.
The    scale factor $R$ is restricted to the domain $R>0$,
so that the 
minisuperspace quantization 
deals only with wave-functions defined on the
half-line $(0,\infty )$. 
It is well-known that in such circumstances
one  has to impose  boundary conditions
on the allowed wave functions otherwise the relevant differential
operators  will not be self-adjoint.
The need to  impose boundary conditions to ensure self-adjointness 
has been  long recognized by practitioners of the 
Arnowitt-Deser-Misner (ADM) reduced phase space formalism as applied to  quantum cosmology 
[8,10-12], and very recently it has also been seen to have non-trivial cosmological
implications in the Wheeler-DeWitt approach [13]. 

In the present case the operator $\hat H$ given by Eq.(3.6) with $k=0$ is self-adjoint in the inner product

$$(\psi ,\phi ) = \int_0^{\infty} R^4 \psi ^*(R) \phi (R) dR \eqno(3.7)$$
\\

\noindent if its domain is suitably specified. 
The operator $\hat H$ is symmetric if

$$(\psi ,{\hat H}\phi )= \int_0^{\infty} \psi ^* (R) \frac{d^2\phi (R)}{dR^2}dR =  
 \int_0^{\infty}  \frac{d^2\psi (R)^*}{dR^2}\phi (R)dR =  
({\hat H}\psi ,\phi )\,\,\, , \eqno(3.8)$$
\\
\noindent and, as in the case of  $ d^2/dR^2$ on $L^2 (0,\infty)$, 
it is well known that the  domain of self-adjointness of
the Hamiltonian operator $ \hat H$  comprises   only those wave 
functions that obey 
 
$$\psi ^{\prime}(0) = \alpha \psi (0) \eqno(3.9)$$
\\
\noindent with $\alpha \in (-\infty ,\infty ]$. 
For the sake of simplicity, here we shall
address ourselves in detail only  to
the cases $\alpha =\infty$ and
$\alpha =0$, that is, the boundary conditions we
shall be mainly concerned with are

$$  \Psi (0,T) = 0 \eqno(3.10a)$$

\noindent or

$$\Psi^{'} (0,T) = 0 \,\,\, , \eqno(3.10b) $$  
\\                                                                   
\noindent where the prime denotes partial derivative with respect to $R$.

Let us look for stationary solutions
to Eq.(3.4), that is, solutions of the form

$$\Psi (R,T) = e^{iET} \psi (R) \,\,\, , \eqno(3.11)$$
\\
\noindent where $E$ is a real parameter.  Then the equation for $\psi (R)$
becomes

$$ \frac{1}{24} \frac{d^2 \psi}{dR^2} + (ER^4 - 6kR^2 )
\, \psi = 0 \,\,\, . \eqno(3.12)$$   
\\
\noindent The above equation coincides with the time-independent Wheeler-DeWitt
equation written by other authors, occasionally
with the help of somewhat obscure methods [14], with
$E$ playing here the role of the cosmological constant $\Lambda$.
It should be emphasized that here this equation has been  derived from a 
well-defined action principle and the
cosmological constant has appeared  dynamically from the vacuum degrees 
of freedom.

In the de Sitter case $(k=0)$ it is easy to show  from the above equation   that
the ``cosmological constant" $E$ is positive. Indeed, multiplying Eq.(3.12)    by
$\psi ^*$ and integrating over the half-line one finds

$$ -\int_0^{\infty} \psi ^* (R) \frac{d^2 \psi (R)}{dR^2} 
\, dR = 24E\int_0^{\infty} R^4 \vert \psi (R) \vert ^2 dR \,\,\, \eqno(3.13)$$
\\
\noindent which, after an integration by parts followed by the use of
(3.9), yields, for $\alpha \geq
0$,

$$E = \frac{1}{24}\frac{\alpha\vert\psi (0)\vert ^2 + \int_0^{\infty}\vert d\psi / dR \vert 
^2 dR }{\int_0^{\infty} R^4 \vert  \psi (R) \vert ^2  dR} > 0 \,\,\, 
\eqno(3.14)$$
\\
\noindent as we wished to prove. It should be clear from the above derivation
that the general boundary condition (3.9) is not sufficient to allow us to 
reach 
the
same conclusion. This special property of  conditions (3.9) with $\alpha\geq 0$
is not present in other minisuperspace models, and seems 
to confer this restricted set of boundary conditions   a 
physically privileged status as compared to the general one 
with arbitrary 
$\alpha$.      

The general solution to Eq.(3.12) with $k=0$ is [15]

$$ \psi_E (R) = \sqrt{R}\, \Bigl [ A\, J_{1/6} (\beta R^3 /3) +
B J_{-1/6} (\beta R^3 /3) \Bigr ] \,\,\, , \eqno(3.15)$$             
\\
\noindent where $J_{\nu}$ denotes a Bessel function of the first kind and 
order $\nu$, $A$ and $B$ are arbitrary constants, and

$$\beta = \sqrt{24 E} \,\,\, . \eqno(3.16)$$
\\

    The usual interpretation of $R^4 \vert \Psi \vert ^2$  as a probability density
implies no correlation between $R$ and $T$. The existence of such solutions to the Wheeler-DeWitt equation 
 is perhaps not surprising since de Sitter's spacetime may be regarded as static 
[16] or self-similar  [17].

It follows from the behavior of Bessel functions for small argument that 
in the case of boundary condition (3.10a) the solution is

$$  \psi^{(a)}_E (R) =  \sqrt{R} \, J_{1/6} (\beta R^3 /3) 
 \,\,\, , \eqno(3.17a)$$             
\\
\noindent whereas in the case of boundary condition (3.10b) the solution is

$$  \psi^{(b)}_E (R) =  \sqrt{R} \, J_{-1/6} (\beta R^3 /3) 
 \,\,\, . \eqno(3.17b)$$             
\\
\noindent
From the asymptotic behavior of Bessel functions for small and large argument
one easily checks  that both solutions are square integrable, but their norm induced by the inner product (3.7) is infinite. Thus, states of the Universe with a definite value of the cosmological constant do not exist. Realizable states can only be constr
ucted by superposition
of solutions to the Wheeler-DeWitt equation with different values of the cosmological constant.

For any two states $\psi_1$ and $\psi_2$ belonging to the domain of the Hamiltonian operator, that is, obeying condition (3.9), one has

$$J_{12}(0) =\frac{i}{2} \Biggl( {\psi_1}^* \frac{\partial{\psi_2}}{\partial R} -
 {\psi_2} \frac{\partial{\psi_1}^*}{\partial R} \Biggr )_{R=0} = 0 \,\,\, .\eqno(3.18)$$
\\
\noindent Therefore, as in other minisuperspace models [12,18], here Vilenkin's  wave function of the universe $\Psi$ is ruled out because it is in conflict with the self-adjointness of the Hamiltonian operator.
Indeed, Vilenkin's tunneling boundary condition [2,5] requires the wave function of the universe $\Psi$ to consist only of outgoing modes at singular boundaries of superspace. In the present context this would amount mathematically to $J_{12}(0)>0$ whenev
er
$\psi_1 = \psi_2 =\Psi$, which is impossible.



\vskip .7cm
\noindent{\bf 4. EVOLUTION OF WAVE PACKETS}                     

The stationary solutions (3.17) have infinite norm and play here a role analogous to that of plane waves in the quantum mechanics of the free particle, that is, finite-norm solutions can be constructed by superposing them. The general solutions to the Whe
eler-DeWitt equation (3.5) with $k=0$  are given by the continuous linear combinations

$$\Psi^{(\sigma )} (R,T) = \int_0^{\infty} c^{(\sigma )}(E) e^{iET} \psi ^{(\sigma )}_E (R)\,\,\,\,\, , \,\,\,\,\, \sigma = a,b \,\,\,\,\, , \eqno(4.1)$$
\\
\noindent where the superscript is used to distinguish the wave functions that obey the boundary condition (3.10a) from those that obey (3.10b). According to the Appendix, the probability distribution of values of the cosmological constant is given by

$$\rho^{(\sigma)} (E) = \frac{1}{4}\, \vert c^{(\sigma)}(E)\vert ^2 \,\,\,  ,\eqno(4.2)$$
\\
\noindent assuming, of course, that $\Psi^{(\sigma )} (R,T)$ is normalized in the inner product (3.7).

We shall consider simple but illustrative examples of wave-packet solutions to the Wheeler-DeWitt
equation obeying each of the boundary conditions (3.10).
Introducing the parameter 

$$\lambda = \frac{\sqrt{24E}}{3}\eqno (4.3)$$ 
\\

\noindent we can write 
(4.1) as

$$\Psi^{(\sigma )} (R,T) = \sqrt{R}\int_0^{\infty} a^{(\sigma )}\,(\lambda ) \, e^{i3{\lambda}^2 T/8}\, J_{\nu}(\lambda R^{3})\, d{\lambda}\eqno(4.4)$$
\\

\noindent where $\nu=+1/6$ or  $\nu=-1/6$ according to whether $\sigma = a$ or $b$,
and

$$a^{(\sigma )}(\lambda)=\frac{3\lambda}{4}\,c^{(\sigma )}(\frac{3{\lambda}^{2}}{8})\, .\eqno(4.5)$$.
\\

\noindent The choice  

$$a^{(\sigma )}(\lambda)={\lambda}^{{\nu}+1} e^{-{\alpha}{\lambda}^{2}}\,\,\,\,\, , \,\, \,\,\, {\alpha}>0 \,\,\,\,\, , \eqno(4.6)$$
\\

\noindent  with $\nu = +1/6$ for $\sigma = a$ and  $\nu = -1/6$ for $\sigma = b$, is particularly simple because it enables us to perform the integration in (4.4) and express the wave function of the Universe in terms of elementary functions [19]. In the 
case $\nu=1/6$
we find

$$\Psi^{(a)} (R,T) = \Bigl[ 2(\alpha - \frac{3iT}{8})\Bigr] ^{-7/6}\, R \, \exp\Bigl(- \frac{R^6}{4(\alpha - \frac{3iT}{8})}\Bigr) \,\,\,\,\, , \eqno(4.7)$$
\\

\noindent whereas for  $\nu=-1/6$ the result is

$$\Psi^{(b)} (R,T) = \Bigl[ 2(\alpha - \frac{3iT}{8})\Bigr ] ^{-5/6}\, \exp\Bigl(- \frac{R^6}{4(\alpha - \frac{3iT}{8})}\Bigr) \,\,\,\,\, . \eqno(4.8)$$
\\

\noindent The  expectation  value of the scale factor is given by

$$\langle R\rangle_T=\frac{\int_{0}^{\infty} R^4 {\Psi}^{*}(R,T)\, R \,{\Psi}(R,T)dR}{\int_{0}^{\infty} R^4 {\Psi}^{*}(R,T){\Psi}(R,T)dR}\,\,\,\,\, , \eqno(4.9)$$
\\

\noindent and its time dependence  reflects the dynamical evolution of the quantized version of de Sitter's cosmological model.
For the two types of boundary conditions of  interest we find, respectively,

$$\langle R\rangle _T^{(a)}=\frac{\Gamma{(\frac{4}{3})}}{\Gamma{(\frac{7}{6})}}\,\, \Bigl[\frac{64{\alpha}^2 + 9T^2}{32\alpha}\Bigr]^{1/6}\,\,\,\,\, , \eqno(4.10)$$
\\ 

\noindent and

$$\langle R\rangle _T^{(b)}=\frac{1}{\Gamma{(\frac{5}{6})}}\,\, \Bigl[\frac{64{\alpha}^2 + 9T^2}{32\alpha}\Bigr]^{1/6}\,\,\,\,\, . \eqno(4.11)$$
\\

\noindent For sufficiently large values of the  time $T$ the expectation  value $\langle R\rangle _T$ grows at the same rate predicted by the classical solution (2.15), that is, the classical regime is attained for asymptotically large times. Quantum effe
cts make themselves felt only for small enough $T$ corresponding to small $R$, as expected.

The dispersion of the wave packets defined by

$$(\Delta R)_T^2 = \langle R^2 \rangle _T -\langle R \rangle _T^2 \eqno(4.12)$$
\\
\noindent is readily computed, with the results

$$ (\Delta R)_T^{(a)} = \Biggl( \frac{\Gamma (3/2)}{\Gamma (7/6)}-\frac{\Gamma (4/3)^2}{\Gamma (7/6)^2}\Biggr)^{1/2}
\Bigl[\frac{64{\alpha}^2 + 9T^2}{32\alpha}\Bigr]^{1/6}\,\,\, ,\eqno (4.13a)$$
\\
$$ (\Delta R)_T^{(b)} = \Biggl( \frac{\Gamma (5/6)\Gamma (7/6)-1}{\Gamma (5/6)^2} \Biggr)^{1/2}
\Bigl[\frac{64{\alpha}^2 + 9T^2}{32\alpha}\Bigr]^{1/6}\,\,\, .\eqno (4.13b)$$
\\
\noindent The wave packets inevitably disperse as time passes, the minimum width being attained at $T=0$. As in the case of the free particle, the more localized the initial state at $T=0$ the more rapidly
the wave packet disperses.

It is important to classify the nature of this model as concerns the presence or
 absence of singularities. For the states (4.7) and (4.8) the expectation value of $R$ never vanishes, showing that these states are nonsingular. The issue of existence or nonexistence of singularities may be addressed from another point of view [20]. We 
can define the probability density 

$$P^{(\sigma )}(R)=R^4\,\, \vert \Psi_E^{(\sigma)}(R)\vert^{2}\,\,\,\,\,\, , \,\,\,\,\, \sigma = a,b \,\,\,\,\, , \eqno(4.14)$$
\\

\noindent for the stationary solutions (3.17a) and (3.17b). The behavior of the Bessel functions for small values of the argument makes it clear that $P^{(\sigma )}(R)\rightarrow 0$ as $R\rightarrow 0$, and thus the singularity is avoided within this mode
l according to this criterion.
Whatever the singularity criterion,  de Sitter's quantum cosmological model is nonsingular just as its classical counterpart.

\vskip .7cm
\noindent{\bf 5. CONCLUSION}

We have shown that taking the vacuum as a perfect fluid with equation of state $p = - \rho$
a Hamiltonian description of de Sitter's cosmological model is possible, which makes subsequent quantization a straightforward process.  This circumvents the problem of insufficient number of degrees of freedom that besets the usual Wheeler-DeWitt quantiz
ation of de Sitter's model. 
The  endowment of the vacuum with  dynamical degress of freedom
makes it possible the introduction of  a time variable which, in turn, 
gives meaning to the dynamical evolution at the quantum level.
The cosmological term is not  
postulated from the beginning, but arises as a manifestation of the vacuum  degrees of freedom.
In our approach states with a definite value of the cosmological constant are ruled out, 
and only those states are realizable that are finite-norm superpositions
of solutions to the Wheeler-DeWitt equation with different values of the cosmological constant.

With  the scale factor  as the sole 
degree of freedom of the gravitational field, stationary  and simple wave-packet solutions to the
Wheeler-DeWitt equation have been found.
It turns out that, for the wave packets
investigated, quantum effects are significant  only for small values of the scale factor, and the classical regime sets in at asymptotically large times. Just like the classical de Sitter model, its quantum counterpart is nonsingular.

\vskip 1cm

\noindent {\bf ACKNOWLEDGMENT}

The authors are grateful to Conselho Nacional de Desenvolvimento Cient\'{\i}fico e Tecnol\'ogico (CNPq), Brazil, for financial support.

\vskip 2cm

\noindent{\bf APPENDIX: COSMOLOGICAL CONSTANT PROBABILITY DENSITY}

Let us start with a normalized state vector $\Psi (R,t)$ for which 
 
$$\Vert \Psi \Vert^2 =\int_0^{\infty}R^4 \vert\Psi (R,t)\vert^2dR =$$

$$\int_0^{\infty}dR R^4\int_0^{\infty}c(E)^* e^{-iEt}{\sqrt R}J_{\nu}(\sqrt{24E}R^3/3)dE\int_0^{\infty}c(E^{\prime}) e^{iE^{\prime}t}{\sqrt R}J_{\nu}(\sqrt{24E^{\prime}}R^3/3)dE^{\prime}\,\,\, .\eqno(A.1)$$
\\
\noindent The change of variables

$$E= \frac{9}{24}\lambda ^2 \,\,\,\,\, , \,\,\,\,\, E^{\prime}= \frac{9}{24}{\lambda^{\prime}} ^2
\,\,\,\,\, , \,\,\,\,\, x=R^3 \,\,\,\,\, , \,\,\,\,\,g(\lambda ) = c(9\lambda^2/24)\exp (-i9\lambda^2 t/24) \,\,\,\,\, ,\eqno(A.2)$$
\\
\noindent leads to

$$\Vert \Psi \Vert^2 =\frac{3}{16} \int_0^{\infty}d\lambda \lambda g(\lambda)^*\int_0^{\infty}d\lambda^{\prime} \lambda^{\prime} g(\lambda^{\prime})
\int_0^{\infty} xJ_{\nu}(\lambda x)J_{\nu}(\lambda^{\prime} x)dx\,\,\, . \eqno(A.3)$$
\\
With the help of Hankel's integral formula [21]

$$f(x) = \int_0^{\infty}\, J_{\nu}(tx)tdt\, \int_0^{\infty }f(\lambda )J_{\nu}(\lambda t)
\lambda d\lambda \,\,\, ,\eqno(A.4)$$
\\
\noindent which is equivalent to the formal equation

$$\int_0^{\infty} xJ_{\nu}(\lambda x)J_{\nu}(\lambda^{\prime} x)dx = \frac{1}{\lambda}\delta (\lambda - \lambda^{\prime})\,\,\, , \eqno(A.5)$$
\\
\noindent one finds

$$\Vert \Psi \Vert^2 =\frac{3}{16}\int_0^{\infty}d\lambda \lambda \vert g(\lambda)\vert^2 = \frac{3}{16}\int_0^{\infty}d\lambda \lambda \vert c(\frac{9\lambda^2}{24})\vert^2 = \frac{1}{4}\int_0^{\infty}\vert c(E)
\vert ^2 dE \,\,\, , \eqno(A.6)$$
\\
\noindent from which Eq.(4.2) follows.
\newpage

\centerline{\bf REFERENCES}
\begin{description}

\item{[1]} B. S. DeWitt, Phys. Rev. {\bf 160}, 1113 (1967).    

\item{[2]} J. A.Halliwell in {\it Quantum Cosmology and Baby Universes}, ed. S.
            Coleman, J. B. Hartle, T. Piran and S. Weinberg (World Scientific,
            Singapore, 1991). This work contains a guide to the literature on 
            quantum cosmology and references to the
            seminal work of Hawking, Hartle, Vilenkin and Linde.               


\item{[3]} E. W. Kolb and M. S. Turner, {\it The  Early Universe}
           (Addison-Wesley, New York, 1994).                      
\item{[4]} J. A. S. Lima and A. Maia, Jr., Phys. Rev. {\bf D52}, 5628 (1995).
\item{[5]} A. Vilenkin, Phys. Rev. {\bf D50}, 2581 (1994), and references therein.   

\item{[6]} B. F. Schutz, Phys. Rev. {\bf D2}, 2762 (1970); {\bf D4}, 3559
            (1971).                                 
\item{[7]} See, for example, S. W. Hawking in {\it Quantum Gravity and
            Cosmology}, ed. H. Sato and T. Inami (World Scientific,
            Singapore, 1986); C. W. Misner, K. S. Thorne and J. A. Wheeler, {\it
            Gravitation} (Freeman, San Francisco, 1973).              

\item{[8]} V. G. Lapchinskii and V. A. Rubakov, Theo. Math. Phys. {\bf 33}, 
           1076 (1977).

\item{[9]} P. Hajicek, Phys. Rev. {\bf D34}, 1040 (1986); see also S. C. Beluardi and                       R. Ferraro, Phys. Rev. {\bf D52}, 1963 (1995). 

\item{[10]} W. F. Blyth and C. J. Isham, Phys. Rev. {\bf D11}, 768 (1975).      

\item{[11]} F. J. Tipler, Phys. Rep. {\bf 137}, 231 (1986).      

\item{[12]} N. A. Lemos, J. Math. Phys. {\bf 37}, 1449 (1996).   

\item{[13]} J. Feinberg and Y. Peleg, Phys. Rev. {\bf D52}, 1988 (1995).   

\item{[14]} M. L. Fil'chenkov, Phys Lett. B354, 208 (1995).
                  
\item{[15]} F. B. Hildebrand, {\it Advanced Calculus for Applications} (Prentice Hall, Englewood             Cliffs, NJ, 1976), sec. 4.10.    

\item{[16]} W. Rindler, {\it Essential Relativity} (Springer, New York, 1977).

\item{[17]} D. W. Sciama, {\it Modern Cosmology} (Cambridge University Press, Cambridge, 1975).

\item{[18]} N. A. Lemos, Phys. Lett. {\bf A 221}, 359 (1996).   
 
\item{[19]} I. S. Gradshteyn and I. M. Ryzhik, {\it Tables of Integrals, Series, and                         Products}, Corrected and Enlarged Edition (Academic, New York, 1980), formula             6.631(4).
\item{[20]} T. Christodoulakis and C. G. Papadopoulos, Phys. Rev. {\bf D38}, 
            1063 (1988).                                                       

\item{[21]} Bateman Manuscript Project, {\it Higher Transcendental Functions}, vol. II, ed. by
            A. Erd\'elyi (McGraw-Hill, New York, 1953), sec. 7.10.5.

\end{description} 
\end{document}